\documentclass[a4paper,11pt]{article}

\usepackage{jcappub}

\begin{document}

\title{\bf $\delta M$ Formalism and Anisotropic Chaotic Inflation Power Spectrum}

\author[a]{Alireza Talebian-Ashkezari,} 
\author[a]{Nahid Ahmadi}

\affiliation[a]{Department of Physics, University of Tehran, \\ Kargar Ave. North, Tehran 14395-547, Iran.}

\emailAdd{atalebian@ut.ac.ir}
\emailAdd{nahmadi@ut.ac.ir}

\date{\today}

\abstract{A new analytical approach to linear perturbations in anisotropic inflation has been introduced in [A. Talebian-Ashkezari, N. Ahmadi and A.A. Abolhasanib, JCAP 03(2018)001] under the name of $\delta M$ formalism. In this paper we apply the mentioned approach to a model of anisotropic inflation driven by a scalar field, coupled to the kinetic term of a vector field with a $U(1)$ symmetry. The $\delta M$ formalism provides an efficient way of computing tensor- tensor, tensor- scalar as well as scalar- scalar 2-point correlations that are needed for the analysis of the observational features of an anisotropic model on the CMB. A comparison between $\delta M$ results and the tedious calculations using in-in formalism shows the aptitude of the $\delta M$ formalism in calculating accurate two point correlation functions between physical modes of the system.}

\maketitle
\section{Introduction}
The main statistical features of cosmic microwave background (CMB) can be explained by an inflationary model. The zeroth order predictions of this model are sufficient for solving the flatness and horizon problems in the context of big bang theory of cosmology \cite{Inflation}. The fluctuations of a single field inflation model are supposed to be statistically isotropic, Gaussian and scale invariant and the overall Planck results \cite{PLANCK} strongly supports these features. However, there are some anomalies in the data \cite{Ade:2015hxq,Schwarz:2015cma} which could not be addressed in the context of simple models of inflation. Those data show that the statistics of the CMB anisotropies does not possess full rotational invariance; this is at odds with mechanisms of generation of perturbations in standard models and suggests inflationary models with an anisotropic stage. To our knowledge, the best model to generate anisotropic perturbations, is to employ gauge fields with a time-dependent gauge kinetic coupling such that the action has the form $\frac{1}{4} f^2(\phi) F_{\mu\nu} F^{\mu\nu}$, in which $\phi$ is the inflation field and $F_{\mu\nu}$ is the $U(1)$ gauge field strength \cite{Watanabe:2009ct,Watanabe:2010fh,Emami:2010rm,Bartolo:2012sd,Ohashi:2013qba,Emami:2013bk,Gumrukcuoglu}. These models first introduced in \cite{Watanabe:2009ct}, are known to be free of ghost instabilities \cite{Himmetoglu:2008hx,Himmetoglu:2008zp,Himmetoglu:2009qi}.

An interesting property of $f^2 F^2$ models is that if one chooses $f(\phi)$ such that $f(\phi) \propto a^2$, in which $a(t)$ is the scale factor, then the inflationary system admits an attractor solution in which metric anisotropy reaches a small but cosmologically detectable level and gauge fluctuations remain scale invariant. In these models, the predictions for curvature perturbations power spectrum \cite{Bartolo:2012sd,Emami:2013bk,Gumrukcuoglu} and tensor sectors in \cite{Gumrukcuoglu,Chen:2014eua} have been studied. These analyses are based on in-in formalism which is technically difficult due to anisotropic background dynamics.

For isotropic inflationary models, there is a technically much easier approach, called $\delta N$ formalism, to handle calculations of curvature perturbations. This formalism has been extended to the anisotropic background models \cite{Abolhasani:2013zya}. In our previous work \cite{Talebian-Ashkezari:2016llx}, we proposed $\delta M$ formalism, which deals with the calculations of non-trivial linear tensor perturbations and their correlations. By (non-)trivial tensor perturbations we mean the the (non-)decaying solutions of perturbated Einstein equations. The non-trivial solutions are sourced by anisotropic energy-momentum tensor, while the trivial tensor modes refer to the homogeneous solutions. It is worthy to note that $\delta M$ formalism is restricted to the perturbations at weak shear limit. 

In this paper we employ this formalism to calculate the power spectra of long wavelength perturbations in above mentioned model of anisotropic inflation in a Bianchi type-${\it I}$ universe filled with the gauge fields. The accuracy of this formalism would be determined by cross-checking it with the established results. We show that our results coincide exactly with those obtained from in-in formalism in \cite{Chen:2014eua}. 

The rest of the paper is organized as follows. A Bianchi type-${\it I}$ universe model which inflates in the presence a $U(1)$ gauge field with the action $f^2 F^2$ is reviewed in Section 2. In Section 3 we briskly review the steps that led to $\delta M$ and $\delta N$ formalisms and then apply the corresponding formulas to the model in hand to find long wavelength metric and matter perturbations, at weak shear limit. The power spectra of curvature and tensor perturbations and the cross-correlations are computed in Section 4. The summary and conclusion are given in Section 5.
\section{Anisotropic Inflationary expansion: Background level}

\subsection{Background geometry}
We assume a homogeneous background which is characterized by a Bianchi type-{\it I} metric
\begin{align}
	\mathrm{d}s^2 = -\mathrm{d}t^2 + e^{2N} \big(e^{2\boldsymbol{M}}\big)_{ij} \mathrm{d}x^i \mathrm{d}x^j.
	\label{metric}
\end{align}
The spatial metic $\gamma_{ij}$ is parametrized as
\begin{equation} \gamma_{ij} =e^{2N}\hat{\gamma}_{ij} \, ,\qquad \hat{\gamma}_{ij}=(e^{2\boldsymbol{M}})_{ij},
\end{equation}
and one can define cosmic expansion rate, $H$ and cosmic shear rate, $\hat{\sigma}_{ij}$ as following
\begin{equation}
	H \equiv\dot{N},~~~~~~~
	\hat{\sigma}_{ij} \equiv \dfrac{1}{2} \dfrac{\mathrm{d}}{\mathrm{dt}} \big( e^{2\mathbf{M}} \big)_{ij} \, .
\end{equation}
The energy momentum tensor is decomposed into energy density $E=T_{00}$, momentum density $J_{i0}=T_{i0}$, isotropic pressure $S = \gamma^{ij} T_{ij}$ and anisotropic stress tensor $\hat{S}_{ij} = e^{-2N} \big( T_{ij} - \dfrac{1}{3} S \gamma_{ij} \big)$. In Bianchi-$I$ models, the momentum constraint is $J_{i0}=0$ and the basic Einstein equations are in the form of 
\begin{eqnarray}
	\dot{N}^2 &=& \dfrac{1}{3M_P^2}E + \dfrac{1}{6} \hat{\sigma}^2 \,,\label{dotN}\\
	\ddot{N} + \dot{N}^2 &=& -\dfrac{1}{3} \hat{\sigma}^2 - \dfrac{1}{6M_P^2} \big( E + S \big),\label{ddotN}\\
	\dot{\hat{\sigma}}^i_j + 3 \dot{N} \hat{\sigma}^i_j-2\hat{\sigma}^{ik}\hat{\sigma}_{kj} &=& \dfrac{1}{M_P^2}\hat{S}^i_j \label{dot_sigma}\,,
\end{eqnarray}
in which $\hat{\sigma}^2 \equiv \hat{\sigma}_{ij} \sigma^{ij}$ is a measure of spatial shear. The energy conservation is also given by
\begin{equation}
	\dot{E} + 3 \dot{N} \big( E + \dfrac{1}{3} S \big) + \hat{\sigma}_{ij} \hat{S}^{ij} = 0 \, \label{energy conservation}.
\end{equation}
We define anisotropic parameter
\begin{equation}
m \equiv \dfrac{\hat{\sigma}}{H} \, \label{m}
\end{equation}
which represents the shear to expansion ratio and characterizes the anisotropy of the expanding universe. In nearly-isotropic Bianchi models, there is a natural splitting between the expansion timescale, $H^{-1}$ and anisotropy timescale $\sigma^{-1}$, such that those correspond to slow and fast time variations, respectively. Therefore, $m$ is a small parameter which may be used as an expansion parameter. Furthermore, we have
\begin{equation}
\frac{\hat{\sigma}^i_j}{H} \sim \mathcal{O}(m)\,,~~~\frac{\hat{S}^i_j}{H^2} \sim \mathcal{O}(m)\,.
\label{near-iso}
\end{equation}
In these models, one can ask two questions: i) How the anisotropies evolve and ii) What are the deviations of above Einstein equations of motion from that of an isotropic FRLW spacetime. At zero-$m$ (or zero shear) limit, above dynamical equations are given by
\begin{eqnarray}
	\dot{N}^2 &=& \dfrac{1}{3M_P^2}E + {\cal O}(m^2) \,, \label{1}\\
	\ddot{N} + \dot{N}^2 &=& - \dfrac{1}{6M_P^2} \big( E + S \big)+ {\cal O}(m^2) \,,\label{2}\\
	\ddot{M}_{ij} + 3 \dot{N} \dot{M}_{ij} &=& \dfrac{1}{M_P^2} \hat{S}^i_j + {\cal O}(m^2)\,.\label{3} \\
	\dot{E} + 3 \dot{N} \big( E + \dfrac{1}{3} S \big) &=&  {\cal O}(m^2) \label{4}\, .
\end{eqnarray}
These equations along with the matter equations form a complete set of equations.{\footnote{We believe that the inclusion of first order corrections (proportional to $m$) is accurate enough for the analysis of the observational features of an anisotropic model on the CMB. The prospect of detecting anisotropies with higher accuracy motivates the development of different methods of computing the perturbations.}} The first question raised above can be answered by solving (\ref{3}). It is easy to show that the solution is given by
\begin{equation}
M_{ij}(t)= M_{ij}(t_0)+\dot{M}_{ij}(t_0)e^{3N(t_0)}\int_{t_0}^{t} \mathrm{d}t' e^{-3N(t')}+\dfrac{1}{M_P^2}\int_{t_0}^{t}\mathrm{d}t' e^{-3N(t')}\int_{t_0}^{t'}\mathrm{d}t'' e^{3N(t'')} \hat{S}^i_j(t'').
\label{sol_M}
\end{equation}
To understand how anisotropies can affect an isotropic background, it is convenient to cast the Einstein equations in the form $R_{\mu\nu}=\frac{1}{M_P^2}\left(T_{\mu\nu}-\frac{1}{2}g_{\mu\nu}T\right)$, and then expand the Ricci tensor to $O\left(m^2\right)$, 
\begin{equation}
R_{\mu\nu}=\bar{R}_{\mu\nu}+R^{\left(1\right)}_{\mu\nu}+R^{\left(2\right)}_{\mu\nu}+...,
\end{equation}
where $\bar{R}_{\mu\nu}$ is constructed with FRLW metric only, $R^{\left(1\right)}_{\mu\nu}$ is linear in $\sigma_{\mu\nu}$ and $R^{\left(2\right)}_{\mu\nu}$ is quadratic in $\sigma_{\mu\nu}$. It is easy to show that the different terms in $\bar{R}_{\mu\nu}$ contain only slow timescales. The crucial observation is that $R^{\left(1\right)}_{\mu\nu}$ contains only fast timescale terms while $R^{\left(2\right)}_{\mu\nu}$ contains both slow and fast timescales, because in a quadratic term a fast timescale term $\sigma_{1}^{-1}$ can combine with $\sigma_{2}^{-1}=-\sigma_{1}^{-1}$ to give a slow varying term. Therefore the Einstein equations can be split into slow and fast parts. Equations (\ref{1}), (\ref{2}) and (\ref{4}) are the slow varying part compared to (\ref{3}) which gives the evolution of the anisotropies.

In the next subsection, we introduce an anisotropic source to appear in the right hand side of (\ref{1}-\ref{4}) and then review the temporal variation of the nearly isotropic Bianchi model at short and long timescales. 
\subsection{Anisotropic inflation due to coupled vector and scalar field}

We consider the following action for the inflaton $\phi$, vector field $A_{\nu}$ and gravitational fields 
\begin{align}
	\mathcal{S} = \int \mathrm{d}^4 x \, \sqrt{-g} \, \bigg[ \dfrac{M^2_{P}}{2}R-\dfrac{1}{2}\partial_{\mu}\phi \partial^{\mu}\bar{\phi} - \dfrac{f^2(\phi)}{4} F_{\mu \nu} F^{\mu \nu} - V(\phi) \bigg].
	\label{action}
\end{align}
Here, $V(\phi)$ is the inflaton potential and $ F_{\mu \nu} = \partial_{\mu} A_{\nu} - \partial_{\nu} A_{\mu}$ is the gauge field strength. The vector field enters only with its kinetic term which is multiplied by a function $f(\phi)$. The gauge field is a coherent space independent field which can affect the background and the scalar field, where the evolution of the latter is given by Klein-Gordon equation 
\begin{equation}
		\partial_{\mu} \Big[ \sqrt{-g} \partial^{\mu} \phi \Big] 
	- \sqrt{-g} \Big( V_{,\phi} + \dfrac{f \, f_{\phi}}{2} F_{\mu\nu}F^{\mu\nu} \Big) = 0 \, . \label{KG2}
\end{equation}
The vector field itself satisfies Maxwell's equations
\begin{align}
	\frac{1}{\sqrt{-g}}\partial_{\mu} \Big( \sqrt{-g} f^2 F^{\mu\nu} \Big) = 0 \,.\label{Maxwell}
\end{align}
The energy momentum tensor for Eistein equation $G_{\mu\nu}=\frac{1}{M_P^2}T_{\mu\nu}$ is given by
\begin{align}
	T_{\mu\nu} =  f^2 F_{\mu\alpha}F^{\alpha}_{\nu} + \partial_{\mu}\phi \partial_{\nu}\phi - g_{\mu\nu} V_{eff} \\
	V_{eff} \equiv \dfrac{f^2}{4} F_{\alpha\beta}F^{\alpha\beta} + \dfrac{1}{2}\partial_{\alpha}\phi \partial^{\alpha}\phi+ V. 
\end{align}
The background gauge field is a coherent homogeneous field; so we have $A_{\mu} = \big( A_0(t),A_i(t) \big), i=1,2,3$. In this model, the local $U\left(1\right)$ gauge invariance of the vector field is preserved and we can use the gauge freedom to constrain $A_{\mu}$. A convenient choice is $A_0=0$. Note that in this model, the momentum density vanishes $T_{i0}= 0$ and the non-zero components of energy momentum tensor are 
\begin{eqnarray}
	E &=& \dfrac{1}{2} f^2 \dot{A}^2 +\left(\dfrac{1}{2} \dot{\phi}^2 + V\right) \,,\label{E}\\
	S &=& \frac{1}{2}f^2 \dot{A}^2 +3 \left(\dfrac{1}{2} \dot{\phi}^2 - V\right) \, \\
	e^{2N} \hat{S}_{ij} &=& -f^2 \big( \dot{A}_i \dot{A}_j -\dfrac{1}{3} \dot{A}^2 \gamma_{ij} \big).
	\label{Shat2}
\end{eqnarray}
Here $A^2 \equiv e^{2N} \hat{\gamma}_{ij} A^i A^j$, and $\dot{A}^2 \equiv e^{2N} \hat{\gamma}_{ij} \dot{A}^i \dot{A}^j$. In the above gauge, the Klein-Gordon equation \eqref{KG2} is given by
\begin{equation}
	\ddot{\phi} + 3 \dot{N} \dot{\phi} + V_{,\phi} - f f_{\phi} \dot{A}^2 = 0.
	\label{KG}
\end{equation}
The Maxwell equation \eqref{Maxwell} can be integrated to give
\begin{equation}
		\dot{A}^i = p^i f^{-2} e^{-3N}\,\label{A^i},
	\end{equation}
	where $p^i$ denotes a constant of integration. We define 
	\begin{eqnarray}
		\dot{A}_i &=& \gamma_{ik} \dot{A}^k = p_k f^{-2} e^{-N}\,,\label{A_i}\\
		\dot{A}^2 &=& \dot{A}^k \dot{A}_k = p^2 f^{-4} e^{-4N}\,.\label{A^2}
	\end{eqnarray}
	Here $p^i = \hat{\gamma}^{ik} p_k$ and $p^2= p_k p^k$ and note that $\dot{A}_i\neq{\partial_t\left(A_i\right)}$. By inserting the gauge field solution in \eqref{1}-\eqref{3}, these equations can be rewritten as
\begin{align}
	\dot{N}^2  &= \dfrac{1}{6M_P^2} f^2 \dot{A}^2  +\dfrac{1}{6} \dot{\phi}^2 + \dfrac{1}{3M_P^2} V + {\cal O}(m^2) \,, \label{11}\\
	\ddot{N} + \dot{N}^2 &= - \dfrac{1}{6M_P^2} f^2 \dot{A}^2 - \dfrac{1}{3M_P^2} \dot{\phi}^2 + \dfrac{1}{3M_P^2} V+ {\cal O}(m^2) \,,\label{22}\\
	\ddot{M}_{ij} + 3 \dot{N} \dot{M}_{ij}  &= -\dfrac{f^2}{M_P^2} \big( \dot{A}^i \dot{A}_j -\dfrac{1}{3} \dot{A}^2 \delta^i_j \big) + {\cal O}(m^2)\label{33}
	\end{align}
It is easy to see that (\ref{KG}) can be deduced after imposing (\ref{A^i}) into (\ref{4}).\\

We are interested in an inflationary background in which vector field is relevant to the inflaton dynamics; nevertheless the slow roll expansion of the universe continues. Furthermore, in a prolonged anisotropic phase, the vev of the vector field evolves slowly during the inflation and a suitable form of $f\left(\phi\right)$ is determined as 
\begin{equation}
f(\phi) =e^{-2cN} =  e^{\frac{2 c}{M_P^2} \int \frac{V}{V_{,\phi}} \mathrm{d}\phi} \,, \label{f}
\end{equation}
where $c$ is a numerical constant. Using the conventional slow roll approximation,
\begin{equation}
 \dot{N}^2=\frac{1}{3M_P^2}V,\label{N-V slow roll}
\end{equation}
 equation (\ref{KG}) can be integrated to give 
\begin{eqnarray}
e^{4N+\frac{4c}{M_P^{2}}\int \frac{V}{V_{,\phi}} \mathrm{d}\phi}\approx \frac{2c^2 p^2}{c-1} \frac{V}{M_P^{2}V_{\phi}^2}+ {\rm const.} e^{-4(c-1)N}\,.\label{c-regimes}
\end{eqnarray}
If $\left(c-1\right)$ is positive (negative), the first (second) term on the right hand side of this equation is dominant. For $c<1$, this gives $N\approx\frac{-1}{M_P^{2}}\int \frac{V}{V_{,\phi}} \mathrm{d}\phi+ {\rm cte}$ which is the conventional slow roll inflation. On the contrary, the condition $c>1$ shows a distinct inflationary phase. To describe this phase more precisely, let us rewrite the klein-Gordon equation as
\begin{equation}
	\ddot{\phi} =- 3 \dot{N} \dot{\phi} - V_{,\phi}\left[1- \frac{c}{\epsilon_V}\frac{f^2 \dot{A}^2}{V}\right], 
	\label{KG1}
\end{equation}
where $\epsilon_V$ is the slow-roll parameter defined by $\epsilon_V=\frac{M_P^2}{2}\left(\frac{V_{,\phi}}{V}\right)^2$. As long as $\frac{f^2 \dot{A}^2}{V}<\frac{\epsilon_V}{c}$, the inflaton field rolls down. Using (\ref{A^2}) and (\ref{c-regimes}) in the regime of interest, $c>1$, we get
\begin{equation}
\frac{f^2\dot{A}^2}{V}=\frac{c-1}{c^2}\epsilon_V.\label{ratio}
\end{equation}
Therefore, if the condition ${f^2\dot{A}^2}/{V}\ll\epsilon_V /c\ll1$ holds initially, the slow roll inflation is realized. In this way, there is an attractor solution in both $c$-regimes. It is easy to see that in attractor regime, the scalar field potential force is balanced with two frictional forces induced by Hubble and gauge field and the terminal velocity is given by $\dfrac{d \phi}{d N} = \left(I-1\right)\frac{V_{\phi}}{V}$, where $I\equiv\frac{c-1}{c}$. This shows that for $c>1$, the inflaton dynamics is about $\frac{1}{c}$ times slower than the conventional result. Let us solve this equation for $N$ in case the inflaton is deriven by the chaotic potential, 
\begin{equation}
V=\frac{1}{2}m^2 \phi^2,~~~~~N-N_e=\frac{1}{4M_p^2}\frac{1}{1-I}\left(\phi_e^2-\phi^2\right).\label{phi-N}
\end{equation}

 We need to look at the anisotropy of the inflationary universe in the regime of interest. We neglect $\ddot{M}_{ij}$ term in (\ref{33}) and then combine the resulting expression with (\ref{N-V slow roll}), it reads
\begin{equation}
\frac{\dot{M}_{ij}}{H}=-\frac{p^i p_j-\frac{1}{3}p^2\delta^i_J}{p^2}\frac{c-1}{c^2}+\mathcal{O}(m^2).
\end{equation}
The right hand side of this equation can be identified as $\frac{\hat{S}^i_j}{H^2}$, which is assumed to be of $\mathcal{O}(m)$. We assume that $\left({p^i p_j-\frac{1}{3}p^2\delta^i_J}\right)/{p^2}\sim\mathcal{O}(1)$, and since $\epsilon_V=c\epsilon_H$, we have
\begin{equation}m\sim I\epsilon_H.\end{equation}
Here $\epsilon_H={-\dot{H}}/{H^2}$. This states that although anisotropies are allowed to grow when the anisotropic part of energy momentum tensor of the model is non-zero, inflation enforces an upper bound on how large they can grow. This result exemplifies the inflationary extended cosmic no-hair theorem discussed in \cite{maleknejad:2012}.\\

We are actually interested in the anisotropy effects on the CMB; so in the next section, we will study the perturbations during the anisotropic inflationary phase. Before this, in the next subsection, we talk about anisotropy components in a chosen basis. 
\subsubsection{Decomposition of anisotropy tensor}	
In an anisotropic background, it is very useful to decompose the traceless spatial quantities of metric and matter (such as $M_{ij}, \hat{\sigma}_{ij}, \hat{S}_{ij},...$) in a local basis adapted to a given direction $k_i$ \cite{Talebian-Ashkezari:2016llx,Pereira:2007yy}. We define a set of local, orthonormal basis $\{ \hat{k}_{i},e^{2}_{i},e^{3}_{i} \}$ which span a constant time hyper-surface. In perturbation level, $\hat{k}^{i}$ corresponds to the normalized momentum of a mode and bases $\{ e^{2}_{i},e^{3}_{i} \}$ span the subspace orthogonal to $\hat{k}^{i}$. By construction, they satisfy the orthonormalisation conditions
\begin{equation}
e^a_i \hat{k}_i \hat{\gamma}^{ij}=0 \,,~~~~~~~ e^a_i e^b_j \hat{\gamma}^{ij}=\delta^{ab}\,.
\label{e^a}
\end{equation}
 Note that the above bases are defined up to a rotation about the $\hat{k}^i$. Any traceless symmetric spatial 3-tensor has 5 independent components which can be readily decomposed by the local basis. In particular, for $M_{ij}$, we have
\begin{align}
M_{ij} &= \dfrac{3}{2}(\hat{k}_{i} \hat{k}_{j}-\dfrac{1}{3}\hat{\gamma}_{ij} )M_{\Vert}
+ 2\sum_{a=2,3} \hat{k}_{(i} e^{a}_{j)}M_{a} + \sum_{\lambda=+, \times} e^{\lambda}_{ij}M_{\lambda} \, .
\label{X_decomposition}
\end{align}
The components, $M_{\Vert},M_{a}$, can be obtained by applying projection along $\hat{k}^{i} \hat{k}^{j}$, $\hat{k}^{(i} e_{a}^{j)}$. Given a symmetric traceless tensor $M_{ij}$, one can find the part transverse to $\hat{k}^{i}$ by applying $\Lambda_{ij,kl}=P_{ik}P_{jl}-\frac{1}{2}P_{ij}P_{kl}$, where $P_{ij}=\delta_{ij}-\hat{k}_{i}\hat{k}_{j}$ is the projector tensor. When $\hat{k}=(1,0,0)$, we have 
\begin{equation}
e^{\lambda}_{ij}M_{\lambda}=\left(\begin{array}{ccc}
0 & 0 & 0 \\
0 & \frac{M_{yy}-M_{zz}}{2} & M_{yz} \\
0 &  M_{yz} &-\frac{M_{yy}-M_{zz}}{2}
\end{array} \right)
\end{equation}
In (\ref{X_decomposition}), $e^{\lambda}_{ij}$ are the polarization tensors defind by $e^{\lambda}_{ij}=\frac{1}{\sqrt{2}}\left[\left(e^{2}_{i}e^{2}_{j}-e^{3}_{i}e^{3}_{j}\right)\delta^{\lambda}_{+}+\left(e^{2}_{i}e^{3}_{j}+e^{3}_{i}e^{2}_{j}\right)\delta^{\lambda}_{\times}\right]$. For the above choice of $\hat{k}$, those are defined by
\begin{equation}
e^{+}_{ij} = \dfrac{1}{\sqrt{2}} \left(
\begin{array}{ccc}
0 & 0 & 0 \\
0 & 1& 0 \\
0 & 0 & -1
\end{array} \right),~~
e^{\times}_{ij} = \dfrac{i}{\sqrt{2}} \left(
\begin{array}{ccc}
0 & 0 & 0 \\
0 & 0 & 1 \\
0 & 1 & 0
\end{array} \right),
\end{equation} 
and $M^{\lambda}$s are given by
\begin{align}
M^+ &= \dfrac{1}{\sqrt{2}} ~ \big( M_{yy}-M_{zz} \big) \,,\label{M^plus}
\\
M^{\times} &=\sqrt{2} ~ M_{yz} \label{M^times}.
\end{align}

By projecting \eqref{3}, on these bases, it can be rewritten as
\begin{equation}
	\ddot{M}_{\chi} + 3 \dot{N} \dot{M}_{\chi} = \hat{S}_{\chi} + {\cal O}(m^2) \,~~~~~\chi=\Vert,2,3,+,\times.\label{chi}
\end{equation}
Let us denote the angle between $\vec{\dot{A}}$ and $\hat{k}$ direction by $\theta$ and define the unit vector 
\begin{equation}
\hat{n}_i\equiv\frac{\dot{A}_i}{\dot{A}}=\left(\cos\theta,\sin\theta\cos\varphi,\sin\theta\sin\varphi\right),\label{n hat}
\end{equation}
then we have $p_i=p \hat{n}_i$. Using (\ref{A^2}), (\ref{f}) and (\ref{ratio}), it is easy to see that $p=\sqrt{I \epsilon_H V e^{4N(1-c)}}$ and anisotropic stress tensor is given by
\begin{eqnarray}
\hat{S}^i_j &=& I \epsilon_H V \big(\hat{n}^i\hat{n}_j-\dfrac{1}{3}\delta^i_j \big) \,.
\label{Shat4} 
\end{eqnarray}
After subatituting this components in \eqref{sol_M} and neglecting the constant and decaying terms at leading order, we obtain
\begin{eqnarray}
M_{ij} = I \epsilon_H N \big(\dfrac{\hat{A}^i\hat{A}_j}{\dot{A}^2}-\dfrac{1}{3}  \delta^i_j \big) \,.
\end{eqnarray}
where $3M_P^2H^2=V$ is used. The $M^\lambda$ solutions can be read easily from \eqref{M^plus} and \eqref{M^times} 
\begin{eqnarray}
M^+ &=& \dfrac{1}{\sqrt{2}} I\epsilon_H N \Big( \dfrac{\dot{A}_z \dot{A}^z - \dot{A}_y \dot{A}^y}{\dot{A}^2} \Big)\,,\label{M_+}\\
M^{\times} &=&\sqrt{2} I\epsilon_H N \Big( \dfrac{\dot{A}_z \dot{A}^y }{\dot{A}^2} \Big)\,.\label{M_times}
\end{eqnarray}

\section{Anisotropic Inflationary expansion: Perturbation level}
Having discussed the anisotropic inflationary model in a Bianchi-{\it I} background universe, it is time to tackle the problem in the presence of the perturbations. We write the perturbed metric in the ADM ( 3+1 decomposition) form
\begin{align}
\mathrm{d}s^2 &= \Big(-\alpha^2 + \beta_k \beta^k \Big) \mathrm{d}t^2 + 2 \beta_k \mathrm{d}x^k ~ \mathrm{d}t ~ + e^{2\mathcal{N}}\Big(e^{2\boldsymbol{\mathcal{M}}}\Big)_{ij} \mathrm{d}x^i \mathrm{d}x^j \, ,
\label{ADMmetric}
\end{align}
where $\alpha(t,\mathbf{x})$ and $\beta^i(t,\mathbf{x})$ are lapse function and shift vector, respectively. $\mathcal{N}(t,\mathbf{x})$ is a scalar function and $\mathcal{M}_{ij}(t,\mathbf{x})$ is a traceless $3 \times 3$ matrix. If we separate the homogeneous and inhomogeneous parts of $\mathcal{N}$ and $\mathcal{M}_{ij}$, we get 
\begin{align}
\mathcal{N}(t,\mathbf{x}) &=N(t)+\psi(t,\mathbf{x}) \, ,\label{N_psi}\\
\mathcal{M}_{ij}(t,\mathbf{x}) &=M_{ij}(t)+h_{ij}(t,\mathbf{x}) \, ,\label{M-h}
\end{align}
in which $\psi$ and $h_{ij}$ are metric perturbations. At linear order, spatial part of the metric can be rewritten as
\begin{equation}
g_{ij} = a^2 \big[(1+2\psi)\delta_{ij}+2M_{ij}+2h_{ij}\big].
\label{g_ij}
\end{equation}
For the matter part (scalar and gauge fields), the perturbations have the following general form
\begin{eqnarray}
\phi (t,\mathbf{x}) &=& \bar{\phi}(t) + \delta \phi (t,\mathbf{x}) \,,\\
A_i (t,\mathbf{x}) &=& \bar{A}_i (t) + \delta A_i(t,\mathbf{x}) \,.
\end{eqnarray}
\subsection{Perturbations solutions}
 The field equations with perturbations have been analyzed in \cite{Talebian-Ashkezari:2016llx}. To establish the long wavelength perturbations in an anisotropic background, a smoothing procedure at the level of the field equations is implemented. It is shown that a good approximation to the perturbed equation of motion associated with (\ref{dot_sigma}) on scales greater than a physical size\footnote{It is good to call it {\it anisotropy} horizon}. $T^{-1}$ is established by the leading term in an expansion in spatial gradient of the inhomogenities with parameter $\epsilon\equiv\frac{k}{T}$. Here, $T^{-1}$ is the comoving scale on which the backgrpund 3-metric evolves,
\begin{equation}\partial_t \hat{\gamma}_{ij}\sim T \hat{\gamma}_{ij},
\end{equation}
 and we have $T^{-1}\sim m\left(aH\right)^{-1}$. This {\it fast} timescale corresponds to the anisotropy timescale, $\sigma^{-1}$,  and is equal to that of non-adiabatic (vector/ tensor) metric perturbations, while for adiabatic ones, the timescale is $\left(aH\right)^{-1}$.\footnote{In \cite{Talebian-Ashkezari:2016llx}, we showed that in order to have a globally valid Bianchi-{I} metric (\ref{metric}) in the limit $\epsilon\rightarrow0$, the condition $\beta_i=O\left(\epsilon\right)$ must be imposed.} If we further apply a gradient expansion with parameter $\epsilon$ to the perturbed equations associated with (\ref{dotN}-\ref{ddotN}) and (\ref{energy conservation}) and keeping the leading term only, the equations of motion governing the long wavelength scalar perturbations can be found. Those are valid on scales greater the anisotropy horizon; needless to say that these scales are greater the Hubble horizon. The main result of \cite{Talebian-Ashkezari:2016llx}, which is somewhat surprising a priori is that conserved quantities like spatial curvature perturbation $\zeta$ and the nonadiabatic tensor perturbation in comoving gauge, $H_\lambda$ are related to the {\it background} geometric quantities. To be more precise,
\begin{align}
\zeta&=\psi|_{\delta \rho=0}=\delta N ,\label{zeta_dN}
\\
H^{\lambda}&=h^{\lambda}|_{\delta \rho=0}=\delta M^{\lambda} \,, \,~~~~~~~\lambda=+,\times\, \label{h_dM}.
\end{align}
Taking the usual view that decaying perturbations are to be ignored, this analysis is valid to all order in perturbation theory for $\zeta$, but (\ref{h_dM}) gives an approximate solution up to $O\left(m^2\right)$. The $\delta$ variations in these equations state that spatial curvature perturbation $\zeta$ and the nonadiabatic tensor perturbation amplitudes, $H_\lambda$ are related to the variations of integrated expansion and shear between two initial flat and final uniform density hypersurfaces. The relation of $\zeta$ with the variations of integrated expansion for $k\ll aH$ is known as $\delta N$ formula. In this point of view, (\ref{zeta_dN}) seems to be trivial but (\ref{h_dM}), which is called $\delta M$ formula, turns out to be a valuable supplementary formula.\\

As an important application of this formula, we consider the perturbations on an anisotropic inflation model introduced in previous section and calculate these observables. 
To keep track of the order of the different terms when the variation is performed, we need to know the behavior of the vector field perturbation, in particular $\frac{\delta \vec{\dot{A}}}{\dot{A}}$ on the uniform density hypersurface. Using the expression given in \cite{Bartolo:2012sd} for this super horizon fluctuation in an isotropic background, we get the corresponding result for anisotropic background as following
\begin{eqnarray}
\label{mode-deltaA}
\frac{\delta \vec{\dot{A}}}{\dot{A}} = \sum_{\lambda=\pm}\vec{e}_{\lambda} \frac{\sqrt{3}H}{\sqrt{2I\epsilon_{H}k^3}M_P}\left(1+O\left(m\right)\right).   \quad \quad k>aH
\end{eqnarray}
In the following, we will see that with regard to the observational constraints, there is no need to know the corrections of order $m$.\\ 

The changes in $N$ between the flat and uniform density timeslices can be found by perturbing (\ref{ratio}) and the second of (\ref{phi-N}),  
\begin{eqnarray}
\delta I &= I \Big( 2\dfrac{\delta \dot{A}}{\dot{A}} - 4 \delta N \Big) \,\label{I_var}\\
\delta N &= -\dfrac{\phi}{2M_P^2} \delta \phi + N \delta I\,.\label{N_var1}
\end{eqnarray}
In these calculations the variation of $V$ and terms of order $NI\delta N$ have been neglected. By combining equatios (\ref{I_var}) and (\ref{N_var1}), we obtain
\begin{equation}
\zeta = \delta N = -\frac{\phi}{2M_P^2} \delta \phi + 2NI \frac{\delta \dot{A}}{\dot{A}}\,.
\label{N_var}
\end{equation}
The second term in right hand side, which is the result of anisotropy, gives a correction of order $N\sqrt I$ to $\zeta$.
\begin{equation}
\zeta = -\frac{\phi}{2M_P^2} \delta \phi\left[1+O\left(N\sqrt I\right)\right].
\end{equation}
As mentioned in \cite{Watanabe:2010fh, Bartolo:2012sd}, to solve the flatness and horizon problems we require $I < 10^{-6}$ and $N\sim 60$; so the correction is small. Nevertheless, with the small anisotropy of the universe, a large statistical anisotropy in the spectrum of the curvature perturbation could be created. \\
 
The variation of $ M^{\lambda}$, can be found from (\ref{M_+}) and (\ref{M_times}), if the changes in $(I \epsilon_H N)$ and $\Big( \dfrac{\dot{A}_i \dot{A}^j }{\dot{A}^2} \Big)$ are known. By using \eqref{I_var} and \eqref{N_var}, we have
$\delta (I \epsilon_H N) = 2I\epsilon_H N \frac{\delta \dot{A}}{\dot{A}}\left(1+O\left( {\sqrt I}\right)\right)$ and from (\ref{n hat}) we get
\begin{eqnarray}
\delta \Big( \frac{\dot{A}_i \dot{A}^j }{\dot{A}^2} \Big) = \hat{n}_i \frac{\delta\dot{A}^j}{\dot{A}}+\hat{n}^j \frac{\delta\dot{A}_i}{\dot{A}}-2\hat{n}_i \hat{n}^j \frac{\delta\dot{A}}{\dot{A}}.
\end{eqnarray}	
Therefore we obtain
\begin{eqnarray}
h^+ &=& \sqrt{2} I\epsilon_H N \sin\theta \big( \sin\varphi\frac{\delta\dot{A}^z}{\dot{A}}-\cos\varphi\dfrac{\delta\dot{A}_y}{\dot{A}} \big) \,,\label{h_+}\\
h^{\times} &=& i \sqrt{2} I\epsilon_H N \sin\theta \big( \cos\varphi\frac{\delta\dot{A}^z}{\dot{A}}+\sin\varphi\dfrac{\delta\dot{A}_y}{\dot{A}} \big).\label{h_times}
\end{eqnarray}

 Equations (\ref{h_+}-\ref{h_times}) and (\ref{N_var}) are the the main result of this section obtained smartly by applying The $\delta M $ formula. In the next section, we confirm these results by calculating the two-point correlations and comparing with the previous predictions of the model elaborated in \cite{Emami:2010rm}.
\section{Anisotropic Correlations}
In order to study observational signatures of these models, we need to know the correlation functions of the perturbations $\zeta,h^{\lambda}$. Equations (\ref{h_+}-\ref{h_times}) and (\ref{N_var}) state that the behaviour of the superhorizon mode functions $\delta \phi$ and $\delta\dot{A}_i$, related to the initial Bunch-Davies vacuum deep inside the horizon, should also be known. We use the following Fourier convention for a typical perturbation $\delta(t,\mathbf{x})$:
\begin{equation}
\delta_{\boldsymbol{k}} = \int \mathrm{d}^3 \mathbf{x} \,\, \delta(t,\mathbf{x}) \, e^{-i \mathbf{k}.\mathbf{x}} \,,
\end{equation}
so that the power spectrum is
\begin{equation}
\langle \delta_{\mathbf{k}} \, \delta_{\mathbf{k}'} \rangle = \left( 2 \pi \right)^3 \delta^3 \left( \mathbf{k} +  \mathbf{k}'\right) \dfrac{2 \pi^2}{k^3} \mathcal{P}_{\delta}(\mathbf{k}) \,.
\end{equation}
Here, $\langle ... \rangle$ defines the ensemble average of the fluctuations and the dimensionless power spectrum $\mathcal{P}_{\delta}$ is chosen such that the variance of $\delta$ is $\langle \delta \, \delta\rangle = \int_{0}^{\infty} \mathcal{P}_{\delta} \, \mathrm{d}\ln k$. 
For $\delta \phi$ in slow-roll inflation, quantum fluctuations of a light scalar field $(m_{\phi} \ll H)$ in quasi-de Sitter space $(H \approx \mathrm{const.})$ scale with the Hubble parameter, $\mathcal{P}_{\phi}=\left({H}/{2 \pi}\right)^2$. And for the power spectrum of the curvature perturbation in the absence of anisotropy, we have
\begin{equation}
\mathcal{P}_{\zeta}^{iso} = \dfrac{1}{2\epsilon_H M_P^2} \bigg( \dfrac{H}{2 \pi} \bigg)^2 \,.
\label{zeta-zeta-iso}
\end{equation}
In addition, in this anisotropic universe model, $\delta\phi$ and $\delta\dot{A}_i$ are mutually uncorrelated, i.e,
\begin{equation}
\langle \delta\phi(\mathbf{k}) \, \delta\dot{A}_i(\mathbf{k'})\rangle = 0 \,.
\label{uncorrelated}
\end{equation}
As shown in \cite{Bartolo:2012sd,Abolhasani:2013zya}, for $f$ given in Eq. (\ref{f}), the mode functions of the rescaled field $f \delta A_{\lambda}$ is the same as that of a massless scalar field in dS space. Using the attractor solution Eq. \eqref{ratio} one can easily show that on super-horizon scales Eq. \eqref{mode-deltaA} holds. Therefore, the related power spectrum is given by
\begin{equation}
\label{modeA}
\langle \frac{\delta \dot{A}_{i}(\mathbf{k})}{ \dot{A}} \frac{\delta \dot{A}_{j}(\mathbf{k'})}{ \dot{A}} \rangle = \left( 2 \pi \right)^3 \delta^3 \left( \mathbf{k} +  \mathbf{k}'\right) \dfrac{2\pi^2}{k^3} \mathcal{P}_{\delta\dot{A}} P_{ij}(\mathbf{k})\,,~~~~~~~\mathcal{P}_{\delta\dot{A}}=  \frac{3}{I\epsilon_{H} M^2_{P}} \bigg( \dfrac{H}{2 \pi} \bigg)^2.
\end{equation}

$\delta M$ formalism gives us the non-trivial part of the tensorial perturbations which have been sourced by matter perturbations, so to establish the total tensorial power spectrum, the isotropic part, $\mathcal{P}_{h}^{iso} \equiv \mathcal{P}_{+}^{iso}+ \mathcal{P}_{\times}^{iso}=\left({8}/{M_P^2} \right)\left({H}/{2 \pi}\right)^2$, must be added to the final results by hand. In the following, the anisotropy corrections to the above expressions will be determined. Before that, we mention two useful relations which can be inferred from (\ref{modeA})  
\begin{eqnarray}
\langle \frac{\delta \dot{A}(\mathbf{k})}{ \dot{A}} \frac{\delta \dot{A}(\mathbf{k'})}{ \dot{A}} \rangle &= &\left( 2 \pi \right)^3 \delta^3 \left( \mathbf{k} +  \mathbf{k}'\right) \dfrac{2\pi^2}{k^3} \big( 1 - (\hat{n}_k \hat{k}^k)^2 \big) \mathcal{P}_{\delta\dot{A}} \,,\\
\langle \frac{\delta \dot{A}_i(\mathbf{k})}{ \dot{A}} \frac{\delta \dot{A}(\mathbf{k'})}{ \dot{A}} \rangle& = &\left( 2 \pi \right)^3 \delta^3 \left( \mathbf{k} +  \mathbf{k}'\right) \dfrac{2\pi^2}{k^3} \big( \hat{n}_i - (\hat{n}_k \hat{k}^k) \hat{k}_i \big) \mathcal{P}_{\delta\dot{A}} \,.
\label{deltaA_i_deltaA}
\end{eqnarray}
Here the identity $\dfrac{\delta\dot{A}}{\dot{A}}=\dfrac{\dot{A}_k}{2\dot{A}} \dfrac{\delta \dot{A}^k}{\dot{A}}+\dfrac{\dot{A}^k}{2\dot{A}} \dfrac{\delta \dot{A}_ k}{\dot{A}}$ has been used. 
\subsection{Scalar-Scalar Correlation}
By using $\mathcal{P}_{\phi}=\left({H}/{2 \pi}\right)^2$, (ref{modeA}) and \eqref{N_var}, we obtain
\begin{eqnarray}
\langle \zeta(\mathbf{k}) \zeta(\mathbf{k'})\rangle &=& \dfrac{\phi^2}{4M_P^4} \langle \delta\phi(\mathbf{k}) \delta\phi(\mathbf{k'})\rangle + 4 I^2 N^2 \langle \frac{\delta \dot{A}(\mathbf{k})}{ \dot{A}} \frac{\delta \dot{A}(\mathbf{k'})}{ \dot{A}} \rangle \nonumber\\
&=& \left( 2 \pi \right)^3 \delta^3 \left( \mathbf{k} +  \mathbf{k}'\right) \dfrac{2\pi^2}{k^3} \dfrac{1}{2\epsilon_H M_P^2} \bigg( \dfrac{H}{2 \pi} \bigg)^2 \big( 1 + 24IN^2\sin^2\theta \big) \nonumber\\
&=& \left( 2 \pi \right)^3 \delta^3 \left( \mathbf{k} +  \mathbf{k}'\right) \dfrac{2\pi^2}{k^3} \mathcal{P}_{\zeta}^{iso} \big( 1 + 24IN^2\sin^2\theta \big)\,.
\label{zeta-zeta}
\end{eqnarray}
where $\mathcal{P}_{\zeta}^{iso}$ is isotropic power spectrum defined in \eqref{zeta-zeta-iso}. This result coincide with the result obtain in \cite{Emami:2010rm} when $\mathbf{e}=0$,
\begin{eqnarray}
\mathcal{P}_{\zeta} = \mathcal{P}_{\zeta}^{iso} \big( 1 + 24IN^2\sin^2\theta \big)\,.
\end{eqnarray}
\subsection{Tensor-Tensor Correlations}
The calculation of anisotropy in tensor power spectra is as easy as above. Let us start with auto-correlations. For $\langle h^+(\mathbf{k}) h^+(\mathbf{k'})\rangle$ one gets
\begin{eqnarray}
	\langle h^+(\mathbf{k}) h^+(\mathbf{k'})\rangle &=& {2} (I\epsilon_H N)^2 \sin^2\theta \Big[ \sin^2\varphi \langle \frac{\delta \dot{A}^z(\mathbf{k})}{ \dot{A}} \frac{\delta \dot{A}^z(\mathbf{k'})}{ \dot{A}} \rangle + \cos^2\varphi \langle \frac{\delta \dot{A}_y(\mathbf{k})}{ \dot{A}} \frac{\delta \dot{A}_y(\mathbf{k'})}{ \dot{A}} \rangle\Big] \nonumber\\
	&=& \left( 2 \pi \right)^3 \delta^3 \left( \mathbf{k} + \mathbf{k}'\right) \frac{2\pi^2}{k^3} \frac{6I\epsilon_H N^2}{M_P^2} \bigg( \dfrac{H}{2 \pi} \bigg)^2 \sin^2\theta\nonumber\\
	&=& \left( 2 \pi \right)^3 \delta^3 \left( \mathbf{k} +  \mathbf{k}'\right) \dfrac{2\pi^2}{k^3} \frac{3}{4}I\epsilon_H N^2 \mathcal{P}_{h}^{iso} \sin^2\theta,
	\label{h+h+}
\end{eqnarray}
and $\langle h^{\times}(\mathbf{k}) h^{*\times}(\mathbf{k'})\rangle$ gives
\begin{eqnarray}
\langle h^{\times}(\mathbf{k}) h^{*\times}(\mathbf{k'})\rangle &=& 2 (I\epsilon_H N)^2\sin^2\theta \Big[ \cos^2\varphi \langle \frac{\delta \dot{A}^z(\mathbf{k})}{ \dot{A}} \frac{\delta \dot{A}^z(\mathbf{k'})}{ \dot{A}} \rangle + \sin^2\varphi \langle \frac{\delta \dot{A}_y(\mathbf{k})}{ \dot{A}} \frac{\delta \dot{A}_y(\mathbf{k'})}{ \dot{A}} \rangle\Big] \bigg\} \nonumber\\
&=& \left( 2 \pi \right)^3 \delta^3 \left( \mathbf{k} + \mathbf{k}'\right) \dfrac{2\pi^2}{k^3} \dfrac{6I\epsilon_H N^2}{M_P^2} \bigg( \dfrac{H}{2 \pi} \bigg)^2 \sin^2\theta \nonumber\\
&=& \left( 2 \pi \right)^3 \delta^3 \left( \mathbf{k} + \mathbf{k}'\right) \dfrac{2\pi^2}{k^3} \dfrac{3}{4}I\epsilon_H N^2 \mathcal{P}_{h}^{iso} \sin^2\theta.
\label{h*h*}
\end{eqnarray}
The total power spectrum $\mathcal{P}_{h}$ takes the form
\begin{eqnarray}
\mathcal{P}_{h} = \mathcal{P}_{h}^{iso}\left(1+\frac{3}{2}I\epsilon_H N^2 \sin^2\theta\right)\,.
\end{eqnarray}
Apart from a factor of four, which is related to the difference in our definition of tensor perturbation with that of \cite{Emami:2010rm}, this expression coincides with their result in $\mathbf{e}=0$ limit. Similarly, our formalism induces the vanishing cross-correlation between two different polarizations 
\begin{eqnarray}
	\langle h^{+}(\mathbf{k}) h^{\times}(\mathbf{k'})\rangle &=& (I\epsilon_H N)^2 \sin^2\theta \sin2\varphi\left( \langle \frac{\delta \dot{A}^z(\mathbf{k})}{ \dot{A}} \frac{\delta \dot{A}^z(\mathbf{k'})}{ \dot{A}} \rangle - \langle \frac{\delta \dot{A}_y(\mathbf{k})}{ \dot{A}} \frac{\delta \dot{A}_y(\mathbf{k'})}{ \dot{A}} \right) \nonumber\\
	&=& 0.
\label{h*h+}
\end{eqnarray}
\subsection{Scalar-Tensor Cross-Correlations}

There would be no cross correlation if anisotropic interactions are absent. Nevertheless, it seems that $\zeta$ does not see $h^{\times}$ and we have
\begin{eqnarray}
\langle \zeta(\mathbf{k}) h^{\times}(\mathbf{k'})\rangle &=& 2\sqrt{2} I^2\epsilon_H N^2 \sin\theta \Big[ \cos\varphi \langle \frac{\delta \dot{A}(\mathbf{k})}{ \dot{A}} \frac{\delta \dot{A}^z(\mathbf{k'})}{ \dot{A}} \rangle - \sin\varphi\langle \frac{\delta \dot{A}(\mathbf{k})}{ \dot{A}} \frac{\delta \dot{A}_y(\mathbf{k'})}{ \dot{A}} \rangle\Big] \nonumber\\
&=& 0.
\end{eqnarray}
The only nonzero cross correlation at linear order is given by
 \begin{eqnarray}
 \langle \zeta(\mathbf{k}) h^+(\mathbf{k'})\rangle &=& {2\sqrt{2}} I^2\epsilon_H N^2 \sin\theta \Big[ \sin\varphi \langle \frac{\delta \dot{A}(\mathbf{k})}{ \dot{A}} \frac{\delta \dot{A}^z(\mathbf{k'})}{ \dot{A}} \rangle + \cos\varphi \langle \frac{\delta \dot{A}(\mathbf{k})}{ \dot{A}} \frac{\delta \dot{A}_y(\mathbf{k'})}{ \dot{A}} \rangle\Big] \nonumber\\
 &=& -\left( 2 \pi \right)^3 \delta^3 \left( \mathbf{k} +  \mathbf{k}'\right) \dfrac{2\pi^2}{k^3} \dfrac{6\sqrt{2}I N^2}{M_P^2} \bigg( \dfrac{H}{2 \pi} \bigg)^2 \sin^2\theta \cos2\varphi. 
 \label{zeta-h+}
\end{eqnarray}
The power spectrum of $\langle\zeta h^+\rangle$  is therefore 
\begin{equation}
\mathcal{P}_{\zeta h_+} = {-12\sqrt{2}I\epsilon_H N^2}\mathcal{P}_{\zeta}^{iso}\sin^2\theta\cos2\varphi.
\end{equation}
As it is elaborated in \cite{Contaldi:2014zua} this negative scalar- tensor cross correlation reduces the impact of the tensor modes on some observable coefficients, $C_{lm,\acute{l}\acute{m}}$. Apart a factor of two and the dependence to azimuthal angle this result conforms to \cite{Emami:2010rm}. The two dimensional rotational symmetry assumption in \cite{Emami:2010rm} explains the $\phi$ independence in their work.\\  
  
\section{summary and conclusion}

In this work we have studied the anisotropic power spectrum in the $f\left(\phi\right)F^2$ model. This model seems a promising avenue by which the the non vanishing anisotropy can be formulated, although temperature
anisotropy data puts stringent bounds on violation of rotational symmetry during inflation in this model \cite{Komatsu, Fujita}. The motivation for such study could have been to reconcile the broken statistical isotropy of the CMB perturbations. However, what provokes us in this study is different. In this paper, we desired the different power spectra to be calculated by a recently introduced formalism, called as $\delta M$ formalism. Specifically, it is a tool for dealing nontrivial interactions between scalar and tensor modes. We showed that how simple and accurate the calculations of the linear perturbations, and consequently the correlation functions, can be by employing the $\delta M$ formalism. To establish the facts of the matter, a comparison with the known results would be necessary.  

In most papers on this subject, the perturbations live in a Bianchi I geometry with residual 2d isotropy. This does not reduce the number of physical modes but reduces the interactions of different modes that enter the cumbersome calculations of 2-point correlation functions through the in-in formalism. We doubted that some implications of the $f^2F^2$ model may be obscured in these symmetric models; so we considered the perturbations on a general Bianchi I geometry. If there was but one of these implications uncovered before, this could not be sound by this generalization. The level of agreement is unprecedented, compared to the simplicity it brings about, except for the dependence on the azimuthal angle, which shows up in scalar-tensor correlation. This can be explained by the fact that scalar perturbations experience the magnitude of the anisotropy only while tensors are affected by the anisotropy components in different directions, as well. The credit of this novel result attributes to $\delta M$ formula.

This study allows us to conclude that powerful $\delta M$ approach leads not only to re-derivation of old and established results, but may contribute to the computation of observables in other anisotropic models.

 \acknowledgments
Authors would like to thank the University of Tehran for supporting this project under the grants provided by the university research council.
 

\end{document}